\documentclass[%
 reprint,
superscriptaddress,
 amsmath,amssymb,
 aps,
pra,
]{revtex4-2}
\usepackage{orcidlink}
\usepackage{graphicx}
\usepackage{braket}
\usepackage{xcolor}
\usepackage{comment}
\usepackage{booktabs}
\usepackage[normalem]{ulem}

\begin{document}

\title{Influence of the Coulomb Interaction on the Radiation of a Few Body System}

\author{Zukhriddin Ruziev \orcidlink{0009-0001-1864-9970}}
\email{zuhriddinrj@gmail.com}
\affiliation{National University of Uzbekistan, Faculty of Physics, University Street 4, Tashkent 100095, Uzbekistan}

\author{Caterina Cocchi \orcidlink{0000-0002-9243-9461}}
\affiliation{Friedrich-Schiller Universit\"at Jena, Institute for Condensed Matter Theory and Optics, 07743 Jena, Germany}
\affiliation{Abbe Center of Photonics, Friedrich-Schiller-Universit\"at Jena, 07745, Jena, Germany}

\author{Anvar Khujakulov \orcidlink{0009-0006-9721-8533}}
\affiliation{Friedrich-Schiller Universit\"at Jena, Institute for Condensed Matter Theory and Optics, 07743 Jena, Germany}

\date{\today}

\begin{abstract}
Understanding energy absorption in laser–plasma interactions is crucial 
for applications ranging from inertial confinement fusion to particle 
acceleration. At the most fundamental level, this energy exchange is 
governed by the motion of individual charged particles in the laser 
field. However, when multiple particles are present, their mutual 
Coulomb interaction can modify the dynamics and, consequently, the 
energy absorption. In this work, we investigate this scenario by 
solving the classical equations of motion for two coupled charged 
particles interacting with an intense laser field. Within the plasma approximation, we find that the Coulomb interaction between the electrons does not significantly affect the energy transfer during the laser–matter interaction. Additionally, we explore the role of the mass asymmetry by considering the classical electron–proton interaction as a target system. Our results highlight the critical importance of the mass: the proton, due to its much slower dynamics, acquires substantially less energy from the field compared to the electron, leading to a pronounced asymmetry in the energy distribution between the two species.

\end{abstract}

\maketitle

\section{Introduction}\label{sec.intro}

An accelerated charged particle radiates electromagnetic energy: this is a
universal process with far-reaching implications across multiple domains
of physics, including beam dynamics in accelerator
facilities~\cite{Wiedemann2015,Grillenberger2021,Adderley2024}, collective plasma processes~\cite{Sagiv2002,Chen2016,Becoulet2026}, relativistic collisions for research beyond the Standard Model~\cite{Ellis2007,Rappoccio2019}, and
high-energy astrophysics~\cite{Rybicki1979,Gaisser1995ParticleAstrophysics,Aharonian2008HighEnergyAstrophysics}. In all such settings, the mutual
electromagnetic interaction between particles and in particular the
Coulomb repulsion between like charges plays a central role in
determining the collective dynamics, requiring a treatment on the same
footing as the external driving field~\cite{Joshi2002}. This is a fundamental aspect of plasma physics, where space-charge effects and collective behavior govern the system's response \cite{Esarey2009LaserDrivenPlasma,Janev2016QuantumCollisionDynamics}.

Particle-in-cell (PIC) simulations are a powerful tool for modeling laser–plasma interactions, enabling the study of plasma and electron beam acceleration driven by laser fields~\cite{Pukhov2003,Sokolov2009Dynamics,Pukhov2016,Lehmann2017,Schluck2017}. For instance, laser irradiation of a solid target at parallel incidence excites a surface plasma wave whose longitudinal field accelerates electrons into a highly collimated, space-charge-dominated beam~\cite{Shen2021}. More recently, the emerging field of laser-plasma-based acceleration of spin-polarized particle beams has attracted significant attention; a comprehensive review discusses the relevant scaling laws for spin-dependent effects, the dominant depolarization mechanisms, and acceleration schemes for polarized leptons, ions, and gamma quanta for applications in high-energy and nuclear physics~\cite{Reichwein2025}.

While PIC simulations capture the collective plasma response, the underlying radiative processes are best understood at the level of a single charged particle. A key effect that becomes increasingly important at ultrahigh laser intensities is the back-reaction of the emitted radiation on the particle's motion—the so-called radiation reaction (RR). When the electron is accelerated by an intense laser field, it emits synchrotron-like radiation, and the associated energy loss modifies its trajectory and energy gain. In laser–plasma interactions, this radiative loss alters the ponderomotive dynamics and can suppress electron trapping in high-intensity regions by effectively expelling the electron from the field~\cite{Ji2014}. Moreover, the emitted radiation imposes fundamental limitations on the maximum achievable energy gain in laser-driven acceleration schemes~\cite{Sprangle1996}.

According to the classical Lawson--Woodward theorem, a charged particle
cannot gain net energy from a plane, transverse, infinitely extended
laser pulse in vacuum, provided the particle initially moves along the
direction of pulse propagation and the pulse carries no longitudinal
electric field component~\cite{WoodwardLawson1948}. In practice, this
symmetry is broken by finite beam-radius effects, transverse field
gradients, or diffraction~\cite{Malka1996,Stupakov2001}. A single electron
loses energy to synchrotron-like radiation during its interaction with
an intense laser field~\cite{leh20201}. 

Numerical simulations have shown that the RR force becomes dynamically
important when the laser intensity exceeds approximately
$5\times10^{22}\,\mathrm{W/cm^2}$~\cite{Zhidkov2002,
Tamburini2010, leh20201}. In the ultrarelativistic regime, the instantaneous radiated power scales as $\gamma^4$, with $\gamma$ denoting the Lorentz factor. The classical motion of a single electron is well described by the equation of motion~\cite{Capdessus2012,Green2015}, while in a crystal, the dynamics involve both electrons and ions~\cite{Lehmann2019}. As a result, radiation reaction plays a dominant role in determining both the particle's trajectory and its energy evolution. The classical
description of RR is most rigorously formulated through the
Lorentz--Abraham--Dirac~(LAD) equation~\cite{dirac1938}, which,
however, admits unphysical runaway solutions. The Landau--Lifshitz~(LL)
approximation~\cite{landau12lehrbuch} resolves this difficulty by
treating RR as a perturbative correction to the Lorentz force,
expressing the RR four-force entirely in terms of first derivatives of
the external field. This approach has been widely adopted in
strong-field studies~\cite{tambu2020, leh20201} and forms the basis of
the present work.

The ongoing development of next-generation laser and accelerator facilities, such as high-power laser systems capable of generating ultra-relativistic electrons with intensities orders of magnitude higher than the aforementioned value, opens new frontiers in strong-field physics. For example, recent advances in extreme light sources, including the European X-ray Free-Electron Laser (XFEL)~\cite{sfa:xfel17_1} and the Linear Coherent Light Source~\cite{gen:christ16}, have enabled radiation intensities exceeding $10^{23}~\mathrm{W/cm^2}$. These facilities will enable precision tests of classical and quantum electrodynamics in regimes previously accessible only in astrophysical environments~\cite{gen:christ16}.

Despite the extensive literature on single-particle radiation reaction,
relatively little attention has been paid to the interplay between RR
and the mutual Coulomb interaction in few-body systems driven by intense
laser fields. In particular, it remains an open question under what
conditions the Coulomb repulsion between like charges, or between
opposite charges, constitutes a non-negligible correction to the
laser-driven dynamics, and whether RR and Coulomb effects are
simultaneously important or act in distinct parameter regimes.

The aim of the present article is to address these questions for two
representative few-body systems: a pair of electrons, and a classical
proton--electron pair. In both cases, we
solve the coupled relativistic equations of motion including the Lorentz
force, the LL radiation-reaction force, and the mutual
Coulomb interaction, and systematically compare simulations in which
these forces are selectively included or excluded. This allows us to
identify the intensity and momentum regimes in which each force
contributes meaningfully to the dynamics.

This paper is organized as follows. In Sec.~\ref{sec.theor}, we present
the theoretical framework, including the covariant equation of motion,
the LL radiation-reaction force, and the Coulomb
interaction (in dimensionless form). The numerical results are presented
in two parts: the interaction of two electrons with a laser field is
discussed in Sec.~\ref{ssec.2elec}, and the analogous problem for a
electron-proton interaction is examined in Sec.~\ref{ssec.classH}.
Section~\ref{sec.concl} summarizes our findings and discusses their
implications.

\section{Theoretical model}\label{sec.theor} 

Consider the covariant formulation of the dynamics of two relativistic electrons (labeled $\alpha = 1, 2$) subjected to an intense electromagnetic wave, with the mutual Coulomb interaction between the electrons included explicitly. The four components $i = 1,2,3,4$ of the four-velocity
$u^i_\alpha = (\gamma_\alpha, \gamma_\alpha \mathbf{v}_\alpha/c)$ and the proper time
element $ds_\alpha$ are defined as $   u^i_\alpha = \frac{dx^i_\alpha}{ds_\alpha}, \qquad
    ds_\alpha = c\,dt\sqrt{1 - \frac{v_\alpha^2}{c^2}} \equiv \frac{c}{\gamma_\alpha}\,dt, $
where $\gamma_\alpha = (1 - v_\alpha^2/c^2)^{-1/2}$ is the Lorentz factor of electron $\alpha$.
The relativistic four-velocity is related to the four-momentum by $p^i_\alpha = mc\,u^i_\alpha$,
where $m$ is the electron rest mass. We use the notation of Landau and Lifshitz~\cite{landau12lehrbuch}
with the metric tensor $g^{ik} = g_{ik}$, whose diagonal elements are $1, -1, -1, -1$. To avoid any complication, it is assumed that 
the radiation of each particle neither influences the external field nor the other particle.

With the electron charge denoted by $-e$, and including the Lorentz force, the RR force $g^i_\alpha$, and the Coulomb interaction force $C^i_\alpha$ exerted by the other electron on particle $\alpha$, the covariant equation of motion can be written as:
\begin{equation}
    mc\,\frac{du^i_\alpha}{ds_\alpha}
    = -\frac{e}{c}\,F^{ik}u_{\alpha k} + g^i_\alpha + C^i_\alpha,
    \label{eq:eom_covariant}
\end{equation}
where $F^{ik}$ is the electromagnetic field tensor of the laser wave.

The Coulomb four-force on electron $\alpha$ arising from electron $\beta \neq \alpha$ is derived from the electrostatic repulsion between the two like charges. In the instantaneous rest frame, the spatial part of $C^i_\alpha$ reduces to the familiar Coulomb form $e^2/R^2_{\alpha\beta}$. In this (non-retarded) laboratory framework, which is consistent with the non-radiative Coulomb-gauge approximation, the force is given by
\begin{equation}
    C^i_\alpha = \frac{e^2}{\bigl(R_{\alpha\beta}+R_0\bigr)^3}
    \left(x^i_\alpha - x^i_\beta\right)
    \qquad (\beta \neq \alpha),
    \label{eq:coulomb_4force}
\end{equation}
where $R_{\alpha\beta} = |\mathbf{R}_\alpha - \mathbf{R}_\beta|$ is the instantaneous
separation between the two electrons, and the cutoff distance $R_0=2r_e$, set equal to twice the classical electron radius $r_e=2.818\times10^{-15}\,\mathrm{m}$, regularizes the singularity of the Coulomb interaction. Throughout this section, $\mathbf{R}_\alpha$ denotes the physical (dimensional) position
vector of electron $\alpha$ in the laboratory frame, with $x^i_\alpha=(ct,\mathbf R_\alpha)$;
the dimensionless position $\mathbf r_\alpha$, related to $\mathbf R_\alpha$ by
Eq.~\eqref{eq:r_dimless} below, is used only from that point onward.
The RR force $g^i_\alpha$ may be formulated in 
%
%
the LL~\cite{landau12lehrbuch} form as
%
\begin{align}
    g^i_\alpha = \frac{2}{3}\frac{e^2}{c}
    \Biggl[
        &-\frac{e}{mc^2}\frac{\partial F^{ik}}{\partial x^l}\,u_{\alpha k}\,u^l_\alpha
        - \frac{e^2}{m^2c^4}\,F^{ik}F_{lk}\,u^l_\alpha \notag \\
        &+ \frac{e^2}{m^2c^4}\,(F_{lk}\,u^k_\alpha)(F^{lm}u_{\alpha m})\,u^i_\alpha
    \Biggr],
    \label{eq:RR_full}
\end{align}
using the antisymmetric nature of the field tensor ($F^{ik} = -F^{ki}$), see Appendix~\ref{sec.appendix}.
In the adopted spin-polarized classical treatment, we can safely neglect the first term of Eq.~\eqref{eq:RR_full}~\cite{landau12lehrbuch,tambu2020}, thereby obtaining the reduced LL radiation reaction force:
\begin{equation}
    g^i_\alpha \approx
    -\frac{2}{3}\frac{e^4}{m^2c^5}\,F^{ik}F_{lk}\,u^l_\alpha
    + \frac{2}{3}\frac{e^4}{m^2c^5}\,
    (F_{lk}\,u^k_\alpha)(F^{lm}u_{\alpha m})\,u^i_\alpha.
    \label{eq:RR_LL}
\end{equation}
Substituting Eqs.~\eqref{eq:RR_LL} and \eqref{eq:coulomb_4force} into
Eq.~\eqref{eq:eom_covariant}, the complete covariant equation of motion for each
electron $\alpha$ becomes:
\begin{align}
    mc\,\frac{du^i_\alpha}{ds_\alpha}
    &= -\frac{e}{c}\,F^{ik}u_{\alpha k}
    - \frac{2}{3}\frac{e^4}{m^2c^5}\,F^{ik}F_{lk}\,u^l_\alpha \notag \\
    &\quad + \frac{2}{3}\frac{e^4}{m^2c^5}\,
    (F_{lk}\,u^k_\alpha)(F^{lm}u_{\alpha m})\,u^i_\alpha\notag\\
    &\quad + \frac{e^2}{(R_{\alpha\beta}+R_0)^3}
    \left(x^i_\alpha - x^i_\beta\right),
    \label{eq:eom_full}
\end{align}
where time is normalized by the inverse wave frequency $\omega^{-1}$~\cite{leh20201}, implying,
$\omega t \to t$, $\mathbf{v}_\alpha$ by $c$, $\mathbf{p}_\alpha$ by $mc$, and $eF_{ik}/(\omega mc) \to F_{ik}$. This allows introducing 
the dimensionless electric field $e\mathbf{E}/(mc\omega) \to \mathbf{E}$ and the dimensionless
magnetic field $e\mathbf{H}/(mc\omega) \to \mathbf{H}$, corresponding to a dimensionless
vector potential $e\mathbf{A}/(mc^2) \to \mathbf{A}$. The spatial
coordinates are normalized correspondingly as
\begin{equation}
    \mathbf{r}_\alpha = \frac{\omega}{c}\,\mathbf{R}_\alpha,
    \qquad
    \rho_0 = \frac{\omega}{c}\,R_0,
    \label{eq:r_dimless}
\end{equation}
so that $r_{\alpha\beta}\equiv|\mathbf r_\alpha-\mathbf r_\beta| = (\omega/c)\,R_{\alpha\beta}$.

In this dimensionless framework, the equation of motion for electron $\alpha$ becomes:
\onecolumngrid
\begin{align}
    \frac{d\mathbf{p}_\alpha}{dt}
    &=
    \underbrace{
    -\bigl(\mathbf{E} + \mathbf{v}_\alpha \times \mathbf{H}\bigr)
    }_{\mathbf{F}_{\rm LL}}
    +
    \underbrace{
    \mathcal{F}\Bigl[
        \mathbf{E}\,(\mathbf{v}_\alpha \cdot \mathbf{E})
        + (\mathbf{E} \times \mathbf{H})
        + \mathbf{H} \times (\mathbf{H} \times \mathbf{v}_\alpha)
        - \gamma_\alpha^2\,\mathbf{v}_\alpha
        \bigl(
            (\mathbf{E} + \mathbf{v}_\alpha \times \mathbf{H})^2
            - (\mathbf{E} \cdot \mathbf{v}_\alpha)^2
        \bigr)
    \Bigr]
    }_{\mathbf{F}_{\rm RR}}
    \notag \\
    &\quad +
    \underbrace{
    \mathcal{C}\,
    \frac{\mathbf{r}_\alpha - \mathbf{r}_\beta}
         {\bigl(r_{\alpha\beta}+\rho_0\bigr)^3}
    }_{\mathbf{F}_{\rm C}},
    \qquad \beta \neq \alpha,
    \label{eq:eom_dimless}
\end{align}
\twocolumngrid
where the first term, $\mathbf{F}_{\mathrm{LL}}$, is the standard Lorentz force. The second term, $\mathbf{F}_{\mathrm{RR}}$, represents the RR force and is multiplied by the dimensionless prefactor
\begin{equation}
\mathcal{F}
=
\frac{2}{3}\frac{e^2\omega}{mc^3},
\label{eq:prefactor}
\end{equation}
which corresponds to the Landau–Lifshitz radiation-reaction force. Finally, the last term, $\mathbf{F}_{\mathrm{C}}$, accounts for the Coulomb repulsion between the two electrons, with the dimensionless Coulomb prefactor
\begin{equation}
\mathcal C
=
\frac{e^2\omega}{mc^3}
=
\frac{3}{2}\mathcal F.
\label{eq:coulomb_prefactor}
\end{equation}
The scaling relation in Eq.~\eqref{eq:coulomb_prefactor} suggests that
electron-electron correlation may become important during recollision events, whereas radiation reaction remains a
small perturbative correction under optical conditions.
Therefore, Eq.~\eqref{eq:eom_dimless} provides a self-consistent
dimensionless model describing the simultaneous action of the external
laser field, radiation losses, and mutual Coulomb repulsion on two
relativistic electrons. To assess the relative importance of the Coulomb and
RR forces in Eq.\eqref{eq:eom_dimless}, the equations of motion are integrated
under three successive approximations: (i)~Lorentz force alone
($\rm F_L$); (ii)~Lorentz force plus Coulomb interaction ($\rm F_L + F_C$);
and (iii)~all three forces including radiation reaction
($\rm F_L + F_C + F_{RR}$).

For the numerical simulations, we consider an electromagnetic
plane wave propagating along the $x$-axis described by the fields:
\begin{equation}\label{eq.EMfiel}
\mathbf{E}
=
\left(
0,\,
E_y,\,
E_z
\right),
\qquad
\mathbf{H}
=
\left(
0,\,
H_y,\,
H_z
\right),
\end{equation}
where 
\begin{align}
    & E_y = E(\eta)\cos\eta \\
    & E_z = E(\eta)\sin\eta
\end{align}
and
\begin{align}
    & H_y = -E(\eta)\sin\eta \\
    & H_z = E(\eta)\cos\eta,
\end{align}
with
\begin{equation}
E(\eta)
=
\sqrt{2}\,a_0
\exp\!\left[
-\left(
\frac{\eta}{\tau}
\right)^2
\right]
\qquad
(\eta=x-t),
\end{equation}
where $\eta$ plays the role of both the pulse-envelope coordinate and the
wave phase, $a_0$ denotes the normalized laser-field amplitude and $\tau$ is the pulse duration. 
Substituting these field components into
Eq.~\eqref{eq:eom_dimless} yields a coupled nonlinear system
of ordinary differential equations for the momentum and position
components of both electrons, which is integrated numerically
using a fourth-order Runge--Kutta scheme.
\\The circularly polarized Gaussian laser pulse propagates along the
positive $x$-axis with wavelength $\lambda = 800\,\mathrm{nm}$ and
dimensionless pulse duration $\tau = 10$, corresponding to a physical
pulse duration of approximately $4.25\,\mathrm{fs}$.
The numerical
integration is performed with a fixed dimensionless time step
$\Delta t = 0.01\,\omega^{-1}$, corresponding to a physical time step
of $\Delta t_{\mathrm{phys}} \approx 0.424\,\mathrm{as}$, over a total
simulation window of $T_{\mathrm{end}} = 200\,\omega^{-1} \approx
84.9\,\mathrm{fs}$, chosen to be sufficiently long to capture the
complete pulse interaction and the subsequent free-propagation dynamics
of both particles after the pulse has passed. In the adopted
dimensionless unit system the wave number reduces to $k_x = 2\pi
c/(\omega\lambda) = 1$, so that the pulse phase is simply $\eta = x -
t$. The Coulomb singularity is regularised by the cutoff $\rho_{0} =
2r_e\,\omega/c \approx 1.88\times10^{-7}$ (dimensionless) for
$\lambda = 800\,\mathrm{nm}$, and a numerical floor $r_{\min} =
10^{-7}$ is additionally imposed on the inter-particle distance
$r_{\alpha\beta}$ at each Runge--Kutta sub-step to prevent
division-by-zero on sub-cutoff scales. The dimensionless
Landau--Lifshitz prefactor evaluates to $\mathcal{F} \approx
1.474\times10^{-8}$ at optical frequencies, and the Coulomb prefactor
is $\mathcal{C} = \tfrac{3}{2}\mathcal{F}$, consistent with
Eqs.~\eqref{eq:prefactor} and~\eqref{eq:coulomb_prefactor}.

\section{Results}\label{sec.results}

\subsection{The interaction electrons}\label{ssec.2elec}

We now present the numerical results obtained by integrating
Eq.~\eqref{eq:eom_dimless} for two electrons initially separated by a dimensionless distance
$d = 0.0003$, corresponding to a physical distance of
$d_{\mathrm{phys}} = d\cdot c/\omega \approx 38.2\,\mathrm{pm}$. The pair is in antiparallel motion relative to the positive $x$-direction, and with an initial momentum of
$p_0 = 0.577\,m_0 c$, corresponding to an initial Lorentz factor
$\gamma_0 = \sqrt{1+p_0^2/m_0^2c^2} \approx 1.155$.
At the given initial momentum, the de~Broglie wavelength of each
electron is $\lambda_{dB} = 2\pi\hbar/p_0 \approx 4.2\,\mathrm{pm}$,
so that $d \approx 9\,\lambda_{dB}$. This choice ensures that the
initial separation is well above the quantum-mechanical scale of the
electron's wave packet, thereby justifying the use of a purely
classical trajectory description and avoiding any artificial
quantum-interference effects between the two electrons.

\begin{figure*}[t]
    \centering
    \includegraphics[width=\textwidth]{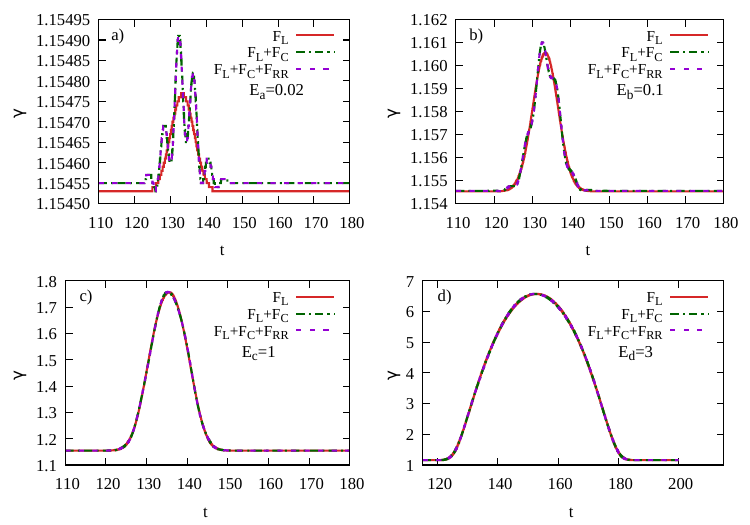}
    \caption{Time evolution of the Lorentz factor $\gamma(t)$ of one
    electron for four peak laser intensities:
    (a)~$I_a \approx 8.58\times10^{14}\,\mathrm{W/cm^2}$ ($E_a=0.02$),
    (b)~$I_b \approx 2.15\times10^{16}\,\mathrm{W/cm^2}$ ($E_b=0.1$),
    (c)~$I_c \approx 2.15\times10^{18}\,\mathrm{W/cm^2}$ ($E_c=1$),
    (d)~$I_d \approx 1.93\times10^{19}\,\mathrm{W/cm^2}$ ($E_d=3$).
    In each panel, three cases are shown: Lorentz force only ($\rm F_L$,
    solid red), with Coulomb interaction ($\rm F_L+F_C$, dash-dotted green),
    and with all forces including radiation reaction
    ($\rm F_L+F_C+F_{RR}$, dashed purple).
    Initial conditions: electron separation $d=0.0003$ (dimensionless),
    initial momentum $p_0=0.577\,m_0 c$ ($\gamma_0 \approx 1.155$),
    laser wavelength $\lambda=800\,\mathrm{nm}$, pulse duration
    $\tau=10$ ($\approx 4.25\,\mathrm{fs}$).}
    \label{fig:gamma_vs_t}
\end{figure*}

We investigate the time evolution of the Lorentz
factor $\gamma(t)$ of one of the two electrons for four values of the
peak laser intensity covering a range from the sub-relativistic to
the highly relativistic regime (Fig.~\ref{fig:gamma_vs_t}). The peak intensity is related to the
dimensionless field amplitude by
$I = (c\varepsilon_0/2)\,(m\omega c/e)^2\,E^2
\approx 2.14\times10^{18}\,E^2\;\mathrm{W/cm^2}$
for $\lambda = 800\,\mathrm{nm}$.
Under a weak peak intensity
$I_a \approx 8.58 \times 10^{14}\,\mathrm{W/cm^2}$ ($E_a = 0.02$),
the laser pulse induces only a very modest excursion of $\gamma$,
which rises from its initial value $\gamma_0 \approx 1.155$ to a maximum
of $\approx 1.155$, corresponding to a relative increase of less than $0.04\%$ (in Fig.\ref{fig:gamma_vs_t}(a)).
In this sub-relativistic regime, the Coulomb interaction between the two
electrons produces a visible splitting of the trajectories, indicating
that at low intensities, when the laser field is insufficiently strong
to dominate the inter-electron dynamics, the Coulomb repulsion plays a
non-negligible role. The radiation-reaction correction remains
indistinguishably small on the scale of panel~(a), confirming that the
Coulomb repulsion between the two electrons constitutes the dominant
correction to single-particle motion while $\rm F_{RR}$ is negligible at this intensity. 
Increasing intensities 
($I_b \approx 2.15 \times 10^{16}\,\mathrm{W/cm^2}$
with $E_b = 0.1$), the laser pulse drives $\gamma$ to a maximum of
$\approx 1.161$, giving rise to a relative increase of roughly $0.6\%$ (in Fig.\ref{fig:gamma_vs_t}(b)).
The Coulomb force
continues to shift the $\gamma(t)$ trajectory relative to the
pure-Lorentz result, while the RR correction is still too small to be
resolved. These results indicate that even in this regime of intense radiation, the Coulomb interaction remains the dominant
correction to the single-particle dynamics.

Increasing the intensity to $I_c \approx 2.15 \times 10^{18}\,\mathrm{W/cm^2}$
($E_c = 1$) marks the entry into the relativistic regime (in Fig.\ref{fig:gamma_vs_t}(c)). Under these conditions, the Lorentz
factor reaches a maximum of $\gamma \approx 1.77$.
In sharp contrast to the lower-intensity cases, the three curves overlap completely, indicating a clear dominance of the laser over both the Coulomb interaction and the RR.
Results obtained at $I_d \approx 1.93 \times 10^{19}\,\mathrm{W/cm^2}$
($E_d = 3$) remain consistent with this trend (in Fig.\ref{fig:gamma_vs_t}(d)).
The Lorentz factor reaches $\gamma \approx 6.7$, confirming the fully in the
ultrarelativistic regime reached under these conditions. Again, the radiation-reaction force $\rm F_{RR}$ remains below the resolution of
all four panels, consistent with the smallness of the prefactor
$\mathcal{F} \approx 1.47\times10^{-8}$ at optical frequencies
(Eq.~\ref{eq:prefactor}).
Taken together, these findings suggest that the threshold for the onset of Coulomb
screening by the laser field lies between
$I \approx 10^{16}$ and $10^{18}\,\mathrm{W/cm^2}$ for the present
initial conditions, and that RR effects would require either higher
intensities or longer interaction times to become observable.

\begin{figure*}[t]
    \centering
    \includegraphics[width=\textwidth]{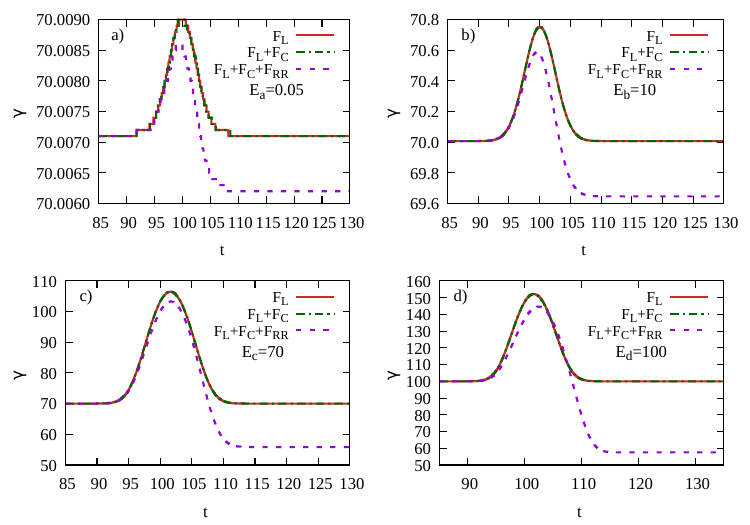}
    \caption{Time evolution of the Lorentz factor $\gamma(t)$ of one
    electron for four peak laser intensities in the ultrarelativistic
    regime ($p_0 \approx 70\,m_0c$, $\gamma_0 \approx 70.007$,
    $d = 3\times10^{-6}$, $d_{\mathrm{phys}}\approx 0.382\,\mathrm{pm}
    \approx 11\,\lambda_{dB}$):
    (a)~$I_a \approx 5.35\times10^{15}\,\mathrm{W/cm^2}$ ($E_a=0.05$),
    The step-like features in the curve are numerical artifacts arising from the finite integration time step, rather than physical effects. 
    (b)~$I_b \approx 2.14\times10^{20}\,\mathrm{W/cm^2}$ ($E_b=10$),
    (c)~$I_c \approx 1.05\times10^{22}\,\mathrm{W/cm^2}$ ($E_c=70$),
    (d)~$I_d \approx 2.14\times10^{22}\,\mathrm{W/cm^2}$ ($E_d=100$).
    Curve notation as in Fig.~\ref{fig:gamma_vs_t}.
    Laser parameters: $\lambda=800\,\mathrm{nm}$, $\tau=10$
    ($\approx 4.25\,\mathrm{fs}$).}
    \label{fig:gamma_vs_t2}
\end{figure*}

To explore the regime where radiation reaction may become significant,
we now consider a qualitatively different scenario: two electrons with
a much larger initial momentum of $p_0 \approx 70\,m_0 c$,
corresponding to $\gamma_0 = \sqrt{1 + p_0^2} \approx 70.007$,
placing the system deeply in the ultrarelativistic regime. The initial separation is
reduced to $d = 3\times10^{-6}$, corresponding to a
physical distance of $d_{\mathrm{phys}} \approx 0.382\,\mathrm{pm}$.
At the corresponding momentum, the de~Broglie wavelength is
$\lambda_{dB} = 2\pi\hbar/p_0 \approx 0.035\,\mathrm{pm}$, so that
$d \approx 11\,\lambda_{dB}$, again safely above the quantum-mechanical
scale and justifying the classical description. Again, we consider four increasing field intensities and explore the behavior of the laser, Coulomb, and RR forces (Figs.~\ref{fig:gamma_vs_t2}) 

At $I_a \approx 5.35 \times 10^{15}\,\mathrm{W/cm^2}$
($E_a = 0.05$), the laser pulse produces only a tiny excursion of
$\gamma$ around its large initial value of $\approx 70.007$ (in Fig.~\ref{fig:gamma_vs_t2}(a)). One caveat is in order regarding the step-like features visible in the curve: these are numerical artifacts stemming from the finite integration time step, and should not be interpreted as physical. This caveat applies throughout the paper whenever such structures appear. In this region, the three
curves $\rm F_L$, $\rm F_L+F_C$, and $\rm F_L+F_C+F_{RR}$ are barely
distinguishable, indicating that under these conditions, neither the Coulomb
interaction nor the RR force produces a significant effect on the electron
dynamics, despite the very small initial separation $d$.
The laser field is still
too weak to significantly perturb the highly relativistic electron, and
the Lorentz factor is dominated by the large initial kinetic energy.

Increasing the intensity to $I_b \approx 2.14 \times 10^{20}\,\mathrm{W/cm^2}$
($E_b = 10$) reveals a markedly different behavior (in Fig.~\ref{fig:gamma_vs_t2}(b)). The coincidence of $\rm F_L$ and $\rm F_L + F_C$ again confirms the negligible contribution of the Coulomb interaction in this setting. In contrast, the inclusion of radiation reaction leads to a pronounced deviation. After the pulse, $\gamma$ is reduced to $\approx 69.6$, corresponding to a net energy loss of about $0.6\%$ that signals a significant radiative
energy loss of the RR force during the interaction. This is consistent with the factor $\mathcal{F}\,\gamma_0^2\,E_b \sim
7\times10^{-4}$, which is large enough to produce a
cumulative effect over the pulse duration.

Shining the system with an intensity $I_c \approx 1.05 \times 10^{22}\,\mathrm{W/cm^2}$
($E_c = 70$) leads to a dramatic radiation-reaction effect (in  Fig.~\ref{fig:gamma_vs_t2}(c)). The RR curve
drops to $\gamma \approx 55$ after the pulse, producing an energy
loss of $\sim21\%$ relative to the initial value. In contrast,
$\rm F_L$ and $\rm F_L+F_C$ remain indistinguishable, confirming
that the Coulomb interaction plays an insignificant role even
at extreme intensities when the electrons are highly relativistic and
the laser field completely dominates the orbital dynamics.
Finally, at the highest intensity
$I_d \approx 2.14 \times 10^{22}\,\mathrm{W/cm^2}$ ($E_d = 100$),
the RR effect becomes most pronounced (see Fig.~\ref{fig:gamma_vs_t2}(d)): the Lorentz factor falls to
$\gamma \approx 55$ after the pulse, with the electron losing a
substantial fraction of its initial energy to radiation. Again, the Coulomb correction remains entirely negligible.

The results shown in Figs.~\ref{fig:gamma_vs_t2} reveal a complementary
picture to those in Figs.~\ref{fig:gamma_vs_t}. In the ultrarelativistic
regime, the Coulomb interaction between the two electrons remains
negligible across the entire intensity range studied, because the large
initial kinetic energy and small inter-electron separation at the moment
of laser interaction render the Coulomb repulsion irrelevant compared to
the laser-driven dynamics. By contrast, the radiation-reaction force
now plays a decisive role at high intensities: it becomes observable
above $I \approx 10^{20}\,\mathrm{W/cm^2}$ and leads to substantial
energy loss at $I \gtrsim 10^{22}\,\mathrm{W/cm^2}$.
The onset of significant RR effects is consistent with the scaling
$\mathcal{F}\,\gamma_0^2\,E \gtrsim 10^{-3}$, which is satisfied for
$E \gtrsim 10$ at $\gamma_0 \approx 70$.
Together, these findings demonstrate
that the importance of Coulomb and radiation-reaction forces is governed
by distinctly different physical conditions: Coulomb effects dominate at
low intensities and low momenta, while RR effects emerge only at
ultrarelativistic momenta and extreme laser intensities.

\begin{figure}[h]
    \centering
    \includegraphics[width=\columnwidth]{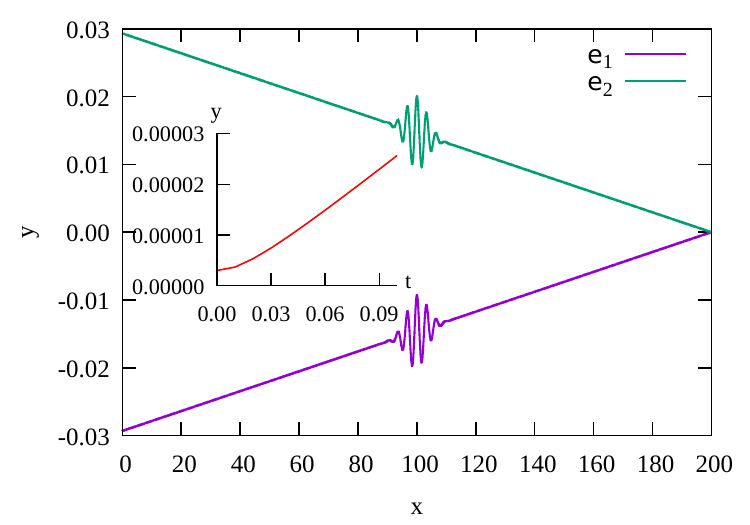}
    \caption{Spatial trajectories of the two electrons $e_1$ (purple)
    and $e_2$ (green) in the $xy$-plane for $p_0 = 70\,m_0c$,
    $E_c = 70$ ($I_c \approx 1.05\times10^{22}\,\mathrm{W/cm^2}$),
    with all forces included ($\rm F_L+F_C+F_{RR}$). Both electrons start
    at $x = 200$ with longitudinal separation $d = 3\times10^{-6}$ and
    travel in the $-x$ direction. Coulomb repulsion deflects $e_1$ and
    $e_2$ symmetrically to $y < 0$ and $y > 0$, respectively. Rapid
    transverse oscillations near $x \approx 100$ mark the interaction
    with the laser pulse. Inset: inter-electron distance $r_{12}(t)$
    at early times ($t \in [0,\,0.1]$), showing the rapid
    Coulomb-driven growth of the separation before the laser arrives.
    All quantities are dimensionless; laser parameters as in
    Figs.~\ref{fig:gamma_vs_t}.}
    \label{fig:trajectory}
\end{figure}

To complement the energy analysis, we display in Fig.~\ref{fig:trajectory} the
spatial trajectories of both electrons in the $xy$-plane for
$p_0 = 70\,m_0c$, $E_c = 70$
($I_c \approx 1.05\times10^{22}\,\mathrm{W/cm^2}$), with all three
forces included ($\rm F_L + F_C + F_{RR}$). 
As investigated in \cite{Feng2005ElectronTrajectory,leh20201}, 
the electron trajectories reproduce the known result: when solved in 
the laser field alone, the motion in the polarization plane becomes 
linear and circular, consistent with the polarization of the laser. 
When Eq.~\eqref{eq:eom_dimless} is solved with the pure Lorentz force 
alone, the rotation radius in the polarization plane is given by
\begin{equation}\label{eq:rotation_radius}
R = \frac{A_0}{\sqrt{1+\langle \mathbf{A} \rangle^2}},
\end{equation}
as shown in \cite{Sarachik1970ClassicalScattering,Feng2005ElectronTrajectory}. 
In this expression, $A_0$ is the dimensionless amplitude, and 
$\langle \mathbf{A} \rangle$ denotes the period-averaged vector potential 
of the pulse from Eqs.\eqref{eq.EMfiel}.

Here, however, the focus is on the Coulomb interaction between the 
two electrons.
The two electrons,
$e_1$ and $e_2$, are initialized at $x = 200$ with a
longitudinal separation $d = 3\times10^{-6}$ along the $x$-axis and
zero transverse offset. Both electrons travel in the negative
$x$-direction with initial velocity $v_0/c \approx 0.9999$, covering a
physical distance of $\approx 25.5\,\mathrm{\mu m}$ over
$\approx 85\,\mathrm{fs}$.
During propagation, the mutual Coulomb repulsion deflects them
symmetrically in the transverse direction: $e_1$ ($e_2$) is displaced toward negative (positive) values of $y$. This Coulomb-driven
transverse separation grows continuously during the free-propagation
phase, reaching a maximum transverse displacement of $|y| \approx 0.03$
($\approx 3.8\,\mathrm{nm}$) in the laser-free region. 

The pulse, propagating in the positive $x$-direction, intercepts
the electrons at $t \approx 42.5\,\mathrm{fs}$ near $x \approx 100$,
where rapid transverse oscillations, superimposed to both
trajectories, reflect the quiver motion driven by
the transverse electric and magnetic fields of the circularly polarized
pulse, with a spatial period of $\approx 2\pi$. After 
interacting with the external field, the trajectories continue their transverse drift,
only slightly modified by the net momentum transfer and the radiative
energy loss established in Figs.~\ref{fig:gamma_vs_t2}.
The inset of Fig.~\ref{fig:trajectory} shows the inter-electron
distance $r_{12}(t)$ during the earliest phase of the motion,
$t \in [0,\,0.1]$, long before the laser pulse arrives. Starting from
the initial separation $r_{12}(0) = d = 3\times10^{-6}$
($\approx 0.382\,\mathrm{pm}$), the distance grows monotonically and
rapidly due to the Coulomb repulsion between the two like charges at
sub-picometer range. This early-time behaviour confirms that the Coulomb
force is indeed active and drives a significant transverse separation
before the laser interaction begins. However, as demonstrated by
Figs.~\ref{fig:gamma_vs_t2}, the resulting change in $\gamma(t)$ remains
negligible throughout, confirming that the Coulomb interaction does not
contribute meaningfully to the energy dynamics in this ultrarelativistic
regime.

\subsection{Electron–proton interaction in a laser field}\label{ssec.classH}

We now turn to the second scenario: a classical electron--proton pair
interacting with the same circularly
polarized Gaussian laser pulse. The proton (charge $+e$, mass
$M_p = 1836.15\,m_e$) is initially held at rest, while the electron
(charge $-e$, mass $m_e$, initial momentum $p_0 = 0.577\,m_0c$,
$\gamma_0 \approx 1.155$) travels in the negative $x$-direction and
encounters both the proton and the laser pulse at $x \approx 133$,
$t \approx 133$. The key distinction from the
two-electron case is that the Coulomb interaction is now
\emph{attractive}, binding the electron to the proton, and that the
large mass ratio $M_p/m_e = 1836.15$ renders the proton nearly
stationary throughout the interaction.
\begin{figure*}
    \centering
    \includegraphics[width=\textwidth]{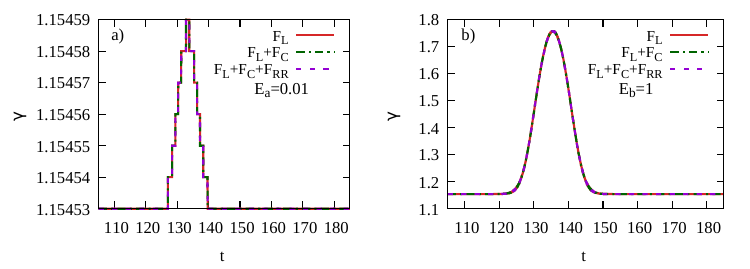}
    \caption{Time evolution of the Lorentz factor $\gamma(t)$ of the
     electron--proton pair ($p_0 = 0.577\,m_0c$,
    $\gamma_0\approx1.155$, initial separation $d = a_0/10 \approx
    5.30\,\mathrm{pm}$) for two peak laser intensities:
    (a)~$I_a \approx 2.14\times10^{14}\,\mathrm{W/cm^2}$ ($E_a=0.01$,
    below ionization threshold);
    (b)~$I_b \approx 2.14\times10^{18}\,\mathrm{W/cm^2}$ ($E_b=1$,
    above ionization threshold).
    Curve notation as in Fig.~\ref{fig:gamma_vs_t}.
    Laser parameters: $\lambda=800\,\mathrm{nm}$, $\tau=10$
    ($\approx 4.25\,\mathrm{fs}$).}
    \label{fig:hydrogen_lowp}
\end{figure*}
Two initial electron--proton separations are considered along
the $y$-axis:
(i)~$d = a_0/10 \approx 4.2\times10^{-5}$ (Bohr radius$a_0 = 5.292\times10^{-11}\,\mathrm{m}$),
corresponding to a physical distance of $d_{\mathrm{phys}} \approx
5.30\,\mathrm{pm}$ and representing the low-momentum case ($p_0 =
0.577\,m_0c$), and
(ii)~$d = a_0 \approx 4.16\times10^{-4}$,
corresponding to $d_{\mathrm{phys}} \approx 52.97\,\mathrm{pm}$,
showcasing the ultrarelativistic regime ($p_0 = 70\,m_0c$).
It is worth noting that for $p_0 = 0.577\,m_0c$, the de~Broglie wavelength of the
electron is $\lambda_{dB} \approx 3.30\times10^{-5}$,
implying $d/\lambda_{dB} \approx 1.26$. This nearly unitary ratio formally places the simulation at the boundary of 
classical description. Nevertheless, the choice $d = a_0/10$ is
physically motivated, as it amplifies the Coulomb force by a factor of
100 compared to the Bohr radius separation, thereby providing the most
stringent test of whether the attractive Coulomb interaction can
produce any observable effect on the electron dynamics. Since no such
effect is found even at this separation (see below), the conclusion
extends \textit{a fortiori} to larger, more classically justified
separations.

The behavior of $\gamma(t)$ is sensitive to the peak laser intensity (Figs.~\ref{fig:hydrogen_lowp}). At $I_a \approx 2.14\times10^{14}\,\mathrm{W/cm^2}$
($E_a = 0.01$), the laser pulse is well below the classical
over-the-barrier ionization threshold
$E_{\mathrm{ion}} \approx 0.128$ (dimensionless), corresponding to
$I_{\mathrm{ion}} \approx 3.52\times10^{15}\,\mathrm{W/cm^2}$.
The electron remains bound to the proton, and $\gamma$ exhibits only a
tiny oscillatory excursion around its initial value of $\approx 1.155$,
with a maximum variation of $\Delta\gamma \approx 6\times10^{-5}$ (in Fig.~\ref{fig:hydrogen_lowp}(a)).
The step-like profile of $\gamma(t)$ over a small absolute range reflects the
quiver motion of the bound electron in the laser field, modulated by
the attractive Coulomb restoring force of the proton. Despite the
strongly enhanced Coulomb force at this sub-bohr separation
(estimated as ${\rm F_C} \sim \mathcal{C}/d^2 \approx 12.8$, which
exceeds the Lorentz force by a factor of $\sim\!10^3$ at $E_a = 0.01$) all three curves perfectly overlap. This result
indicates that the attractive Coulomb binding, while dominant over the
laser force, does not produce a net differential effect on
$\gamma(t)$: it merely confines the electron without 
noticeably altering the laser-induced energy gain.
Increasing the to $I_b \approx 2.14\times10^{18}\,\mathrm{W/cm^2}$ ($E_b = 1$), i.e.,  nearly an order
of magnitude above the ionization threshold, the Lorentz factor reaches a maximum of $\gamma \approx 1.77$, which is identical
to the single-electron and two-electron results at the same intensity
(Fig.~\ref{fig:gamma_vs_t}c). This confirms that the laser field 
completely dominates and the electron behaves essentially as a free particle, while the RR correction remains negligible.

\begin{figure*}
    \centering
    \includegraphics[width=\textwidth]{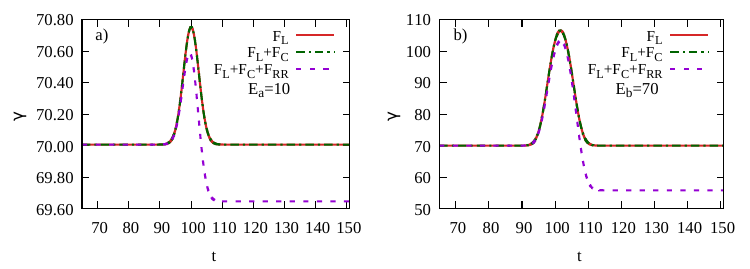}
    \caption{Time evolution of the Lorentz factor $\gamma(t)$ of the
     electron--proton pair ($p_0 = 70\,m_0c$,
    $\gamma_0\approx70.007$, initial separation $d = a_0 \approx
    52.97\,\mathrm{pm}$) for two peak laser intensities:
    (a)~$I_a \approx 2.14\times10^{20}\,\mathrm{W/cm^2}$ ($E_a=10$);
    (b)~$I_b \approx 1.05\times10^{22}\,\mathrm{W/cm^2}$ ($E_b=70$).
    Curve notation as in Fig.~\ref{fig:gamma_vs_t}.
    Laser parameters: $\lambda=800\,\mathrm{nm}$, $\tau=10$
    ($\approx 4.25\,\mathrm{fs}$).}
    \label{fig:hydrogen_highp}
\end{figure*}

We next consider the ultrarelativistic case with $p_0 = 70\,m_0c$
($\gamma_0 \approx 70.007$) and initial separation $d = a_0$,
corresponding to $d_{\mathrm{phys}} \approx 52.97\,\mathrm{pm} \approx
1528\,\lambda_{dB}$, well within the classical regime.
The electron starts at $x = 200$, and the proton is placed at
$x \approx 100$, $y = a_0$, so that the laser pulse, the electron,
and the proton all meet near $x \approx 100$, $t \approx 100$.
At $I_a \approx 2.14\times10^{20}\,\mathrm{W/cm^2}$
($E_a = 10$), the three curves $\rm F_L$, $\rm F_L+F_C$, and
$\rm F_L+F_C+F_{RR}$ nearly coincide before the pulse (in Fig.~\ref{fig:hydrogen_highp}(a)), confirming
that the Coulomb attraction between the electron and proton
plays no significant role in the energy dynamics. After the laser
pulse, however, the $\rm F_L+F_C+F_{RR}$ curve drops sharply to
$\gamma \approx 69.6$, experiencing a net energy loss of $\sim\!0.6\%$, while
$\rm F_L$ and $\rm F_L+F_C$ return to $\gamma_0$. This is precisely the
RR signature already identified in the two-electron
case (Fig.~\ref{fig:gamma_vs_t2}(b)), which demonstrates that the
RR behavior is insensitive to the sign of the inter-particle Coulomb
interaction.

For $I_b \approx 1.05\times10^{22}\,\mathrm{W/cm^2}$
($E_b = 70$), the RR effect is dramatically amplified: $\gamma$ drops
to $\approx 55$ after the pulse, representing an energy loss of
$\sim\!21\%$ (in Fig.~\ref{fig:hydrogen_highp}(b)). The $\rm F_L$ and $\rm F_L+F_C$ curves again coincide
throughout, confirming the irrelevance of the Coulomb force in this setting. These results are quantitatively consistent with the
two-electron case (Fig.~\ref{fig:gamma_vs_t2}c), confirming
that at ultrarelativistic intensities the radiation-reaction energy
loss is a single-particle effect governed by $\mathcal{F}\gamma_0^2 E$
and is independent of whether the companion particle carries the same
or opposite charge.

\begin{figure}
    \centering
    \includegraphics[width=\columnwidth]{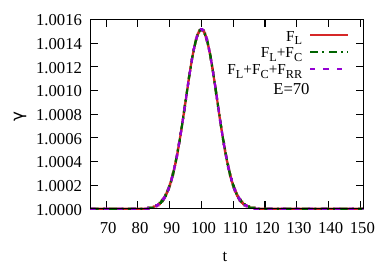}
    \caption{Time evolution of the Lorentz factor $\gamma_p(t)$ of the
    \emph{proton} for $p_0^{\mathrm{electron}} = 70\,m_0c$,
    $E = 70$ ($I \approx 1.05\times10^{22}\,\mathrm{W/cm^2}$),
    with all forces included ($\rm F_L+F_C+F_{RR}$). The proton's energy
    gain is less than $0.2\%$, confirming its near-stationary behavior.
    Laser parameters as in Fig.~\ref{fig:gamma_vs_t}.}
    \label{fig:proton_gamma}
\end{figure}

The Lorentz factor of the \emph{proton} under the same conditions as
in Fig.~\ref{fig:hydrogen_highp}(b) ($p_0^{\mathrm{electron}}
= 70\,m_0c$, $E = 70$,
$I \approx 1.05\times10^{22}\,\mathrm{W/cm^2}$) rises from unity to a maximum of only $\gamma_p \approx
1.0016$, reflecting an energy gain smaller than $0.2\%$ (Fig.\ref{fig:proton_gamma}). All three
curves coincide perfectly, confirming that neither the Coulomb
interaction nor the RR produces any resolvable effect
on the proton dynamics. This result is physically transparent: the
Lorentz force on the proton is scaled by $1/M_p$ relative to that on
the electron, so the proton's response to the laser field is suppressed
by the large mass ratio $M_p/m_e = 1836.15$. Similarly, the RR force
on the proton scales as $(m_e/M_p)^3\sim 1.6\times10^{-10}$, making
it entirely negligible. The proton thus remains effectively stationary
throughout the interaction, fully justifying the heavy-particle
approximation adopted in Sec.~\ref{sec.theor}.

\begin{figure}[h]
    \centering
    \includegraphics[width=0.45\textwidth]{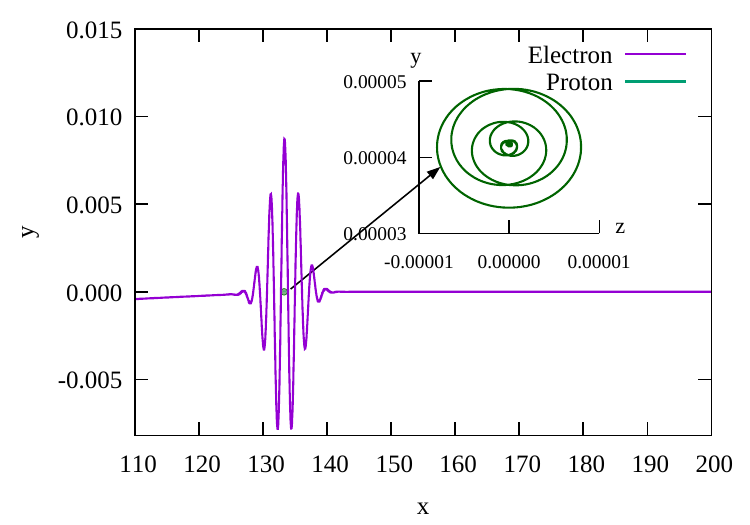}
    \caption{Spatial trajectories of the electron (solid line) and proton
    (dashed line) in the $xy$-plane for $p_0 = 0.577\,m_0c$,
    $E_a = 0.01$ ($I_a \approx 2.14 \times 10^{14}$~W/cm$^2$,
    below ionization threshold), with all forces included
    ($\rm F_L + F_C + F_{RR}$). The electron exhibits rapid quiver oscillations
    near $x \approx 133$, while the proton remains effectively stationary.
    All quantities are dimensionless; laser parameters as in Fig.\ref{fig:gamma_vs_t}.}
    \label{fig:7}
\end{figure}

The spatial trajectory of the electron
in the $xy$-plane for $p_0 = 0.577\,m_0c$ and 
field amplitude $E_a = 0.01$ (below the ionization threshold,
$I_a \approx 2.14 \times 10^{14}$~W/cm$^2$) with all three forces included
($\rm F_L + F_C + F_{RR}$) undergoes rapid transverse
oscillations driven by the laser field, as shown Fig.~\ref{fig:7}. The transverse displacement of the electron
is confined to $|y| \lesssim 0.01$, with no net drift after the pulse,  in contrast to the ultrarelativistic two-electron
case in Fig.~\ref{fig:trajectory}, where a permanent Coulomb-driven transverse separation
develops. This confirms that at sub-relativistic momenta and sub-ionization intensities, the attractive Coulomb binding does not produce any observable symmetry-breaking effect on the trajectories. The corresponding proton trajectory is shown in the inset of Fig.\ref{fig:7}. In contrast, the proton remains nearly stationary over the course of the interaction, in agreement with the heavy-particle approximation. As investigated in \cite{Feng2005ElectronTrajectory}, it has no net 
displacement along the propagation direction ($x$), where it stays at rest. Instead, it oscillates in the $yz$ plane, following the polarization of the laser electric field, which is consistent with previously reported results 
(e.g., Fig.2 in \cite{Feng2005ElectronTrajectory}). The amplitude of these oscillations is about two orders of magnitude smaller than that of the electron. This behavior is expected, as the rotation radius in Eq.\eqref{eq:rotation_radius} 
is inversely proportional to the particle's mass, as discussed in 
Sec.~\ref{sec.theor}.

These results
demonstrate that in the  electron--proton pair, the Coulomb
interaction between the electron and proton remains negligible across
all studied regimes. This finding contrasts sharply with the
two-electron case (Figs.~\ref{fig:gamma_vs_t}a--b), where the
repulsive Coulomb interaction was found to be the dominant correction
at sub-relativistic intensities. The physical reason is clear: in the
 electron--proton pair, the Coulomb force is attractive and acts as a restoring
(binding) force rather than a symmetry-breaking perturbation. While it
confines the electron near the proton, it does not produce a net
differential energy transfer that would distinguish the bound from the
unbound dynamics on the scale of $\gamma(t)$. The radiation-reaction
force, on the other hand, retains its single-particle character and produces
the same energy loss as in the two-electron case whenever
$\mathcal{F}\gamma_0^2 E \gtrsim 10^{-3}$, independently of the sign
or magnitude of the inter-particle Coulomb interaction. 

\begin{table}[h]
\centering
\caption{Role of the Coulomb ($\rm F_C$) and radiation-reaction ($\rm F_{RR}$)
forces relative to the Lorentz force ($\rm F_L$) for the two systems
studied. ``Negligible'': correction unresolvable on the scale of
$\gamma(t)$; ``Observable'': clear deviation visible;
``Dominant'': leading correction to single-particle dynamics.}
\label{tab:summary}
\begin{tabular}{lcccc}
\hline\hline
System & $p_0$ & $I\,[\mathrm{W/cm^2}]$ & $\rm F_C$ & $\rm F_{RR}$ \\
\hline
Two electrons     & $0.577\,m_0c$ & $\lesssim10^{16}$  & Dominant   & Negligible \\
Two electrons     & $0.577\,m_0c$ & $\gtrsim10^{18}$   & Negligible & Negligible \\
Two electrons     & $70\,m_0c$    & $\gtrsim10^{20}$   & Negligible & Observable \\
Two electrons     & $70\,m_0c$    & $\gtrsim10^{22}$   & Negligible & Dominant   \\
H atom (electron) & $0.577\,m_0c$ & all                & Negligible & Negligible \\
H atom (electron) & $70\,m_0c$    & $\gtrsim10^{20}$   & Negligible & Observable \\
H atom (proton)   & $70\,m_0c$    & all                & Negligible & Negligible \\
\hline\hline
\end{tabular}
\end{table}

\section{Conclusions}\label{sec.concl}

We have studied the relativistic laser-driven dynamics of two charged-particle
systems, including a pair of electrons and an electron--proton pair,
treating the Landau--Lifshitz radiation-reaction force and the mutual
Coulomb interaction on equal footing. The central finding,
summarized in Table~\ref{tab:summary}, is that the Coulomb and
radiation-reaction forces act in entirely distinct regimes
and are never simultaneously crucial.
For the two-electron system, the repulsive Coulomb interaction
is dominant at
sub-relativistic intensities
($I \lesssim 10^{16}\,\mathrm{W/cm^2}$, $p_0 = 0.577\,m_0c$),
while radiation reaction becomes the leading effect only at
ultrarelativistic momenta and extreme intensities
($\gamma_0 \approx 70$, $I \gtrsim 10^{20}\,\mathrm{W/cm^2}$),
with its onset governed by the scaling
$\mathcal{F}\gamma_0^2 E \gtrsim 10^{-3}$.
For the  electron--proton pair, the attractive Coulomb force produces
no relevant effect across the entire
intensity and momentum range studied, even at sub-bohr separations
where it exceeds the Lorentz force by three orders of magnitude. In contrast, the radiation-reaction energy loss quantitatively matches the
two-electron result, confirming its single-particle character.
The proton remains effectively stationary throughout the dynamics, validating the heavy-particle approximation.

These results provide practical guidance for strong-field simulations:
inter-particle Coulomb effects should be retained for non-relativistic
electron beams at moderate intensities
($10^{16}$--$10^{18}\,\mathrm{W/cm^2}$), while radiation-reaction
corrections become essential only for highly relativistic particles
($\gamma_0 \gg 1$) at intensities accessible to next-generation laser
facilities such as ELI and the European XFEL~\cite{gen:christ16}.
Extensions to higher intensities where quantum corrections to the
radiation-reaction force become relevant are
left for future work.

\section*{Acknowledgments}
\vspace{-2ex}

Z.R. acknowledges financial support from the Innovative Development Agency of Uzbekistan.

\section*{Data availability}
\vspace{-2ex}

The data that support the findings of this article are openly available in Zenodo: .

\appendix
\section{Lorentz Transformation of the Landau--Lifshitz Radiation-Reaction  Force}
\label{sec.appendix}




One of the essential properties of the relativistic equations, 
they must be the Lorentz invariant. For this purpose, the invariance 
of the LL force is briefly investigated here.

\subsection{ The 4-Vector Form of the Landau--Lifshitz Force}

The LL force given in 4-vector form in Eq.\eqref{eq:RR_full} is

\begin{equation}
\begin{aligned}
g^i = \frac{2}{3}\frac{e^2}{c}
\Biggl[
&-\frac{e}{mc^2}
\frac{\partial F^{ik}}{\partial x^l}
u_{ k}u^l \\
&-\frac{e^2}{m^2c^4}
F^{ik}F_{lk}u^l \\
&+\frac{e^2}{m^2c^4}
(F_{lk}u^k)
(F^{lm}u_{ m})u^i
\Biggr],
\end{aligned}
\label{eq:4force}
\end{equation}

where:
\begin{itemize}
    \item $e$ is the charge of the particle,
    \item $m$ is its mass,
    \item $c$ is the speed of light,
     \item $x^l = (ct,\textbf{r})$ is the 4-position,
    \item $u^i = (\gamma c, \gamma \mathbf{v})$ is the 4-velocity,
    \item $\gamma = 1/\sqrt{1 - \mathrm{v}^2/c^2}$ is the Lorentz factor.
\end{itemize}
The (contravariant form) electromagnetic field tensor $F^{ik}$ and its components are:

\begin{equation}
F^{ik} = 
\begin{pmatrix}
0 & -E_x/c & -E_y/c & -E_z/c \\
E_x/c & 0 & -B_z & B_y \\
E_y/c & B_z & 0 & -B_x \\
E_z/c & -B_y & B_x & 0
\end{pmatrix},
\label{eq:F_tensor}
\end{equation}

with the dual relations:

\begin{align}
F^{0i} &= \frac{E^i}{c}, \quad F^{ij} = -\epsilon^{ijk} B_k, \\
F_{0i} &= -\frac{E^i}{c}, \quad F_{ij} = \epsilon^{ijk} B_k.
\end{align}

Consider a boost along the $x$-axis:

\begin{equation}
\Lambda = 
\begin{pmatrix}
\gamma_0 & -\gamma_0 \beta_0 & 0 & 0 \\
-\gamma_0 \beta_0 & \gamma_0 & 0 & 0 \\
0 & 0 & 1 & 0 \\
0 & 0 & 0 & 1
\end{pmatrix},
\quad \beta_0 = \frac{v_0}{c}.
\label{eq:lambda_matrix}
\end{equation}
where $\gamma_0 = 1/\sqrt{1 - v_0^2/c^2}$.
The 4-vector form Eq.~(\ref{eq:4force}) transforms covariantly:

\begin{equation}
g'^i = \Lambda^i_{\; j} g^j,
\label{eq:4force_transform}
\end{equation}
where $\Lambda^i_{\; j}$ is defined in Eq.\eqref{eq:lambda_matrix}.


\subsection{Lorentz transformation of the terms constituting the LL force}

On the other hand, applying the Lorentz transformation matrix of Eq.~\eqref{eq:lambda_matrix} to the four-vectors and tensors in Eq.~\eqref{eq:4force} should yield transformed components consistent with those presented in Eq.~\eqref{eq:4force_transform}. To demonstrate this, we list each transformed quantity below and explicitly show one component transformation as an illustrative example.

\textbf{a)} The 4-velocity in Eq.\eqref{eq:4force} transforms as a contravariant 4-vector:

\begin{equation}
u^i = (\gamma c, \gamma \mathbf{v})
\label{eq:4velocity_def}
\end{equation}

\begin{equation}
u'^i = \Lambda^i_{\; j} u^j
\label{eq:4velocity_transform}
\end{equation}

When the 0-component is calculated:

\begin{equation}
u'^0 = \Lambda^0_{\; 0} u^0 + \Lambda^0_{\; 1} u^1
\label{eq:time_component_1}
\end{equation}

\begin{equation}
u'^0= \gamma_0 (\gamma c) + (-\gamma_0 \beta_0)(\gamma v_x)
\label{eq:time_component_2}
\end{equation}

\begin{equation}
u'^0 = \gamma_0 \gamma (c - \beta_0 v_x) = \gamma_0 \gamma c \left(1 - \frac{v_0 v_x}{c^2}\right)
\label{eq:time_component_3}
\end{equation}

Since $u'^0 = \gamma' c$, we define:

\begin{equation}
\gamma' = \gamma_0 \gamma \left(1 - \frac{v_0 v_x}{c^2}\right)
\label{eq:gamma_transform}
\end{equation}

The summary of transformed 4-velocity components is:

\begin{eqnarray}
u'^0 &=& \gamma_0 \gamma c \left(1 - \frac{v_0 v_x}{c^2}\right) = \gamma' c, \label{eq:u0_transformed} \\
u'^1 &=& \gamma_0 \gamma (v_x - v_0) = \gamma' v'_x, \label{eq:u1_transformed} \\
u'^2 &=& \gamma v_y = \gamma' v'_y, \label{eq:u2_transformed} \\
u'^3 &=& \gamma v_z = \gamma' v'_z. \label{eq:u3_transformed}
\end{eqnarray}


\textbf{b)} The partial derivative $\partial/\partial x^l$ in Eq.\eqref{eq:4force} transforms as a covariant 4-vector:

\begin{equation}
\frac{\partial}{\partial x'^l} = \Lambda_l^{\; m} \frac{\partial}{\partial x^m}.
\label{eq:partial_transform}
\end{equation}
For the time component, we have:

\begin{equation}
\frac{\partial}{\partial x'^0} = \Lambda_0^{\; 0} \frac{\partial}{\partial x^0} + \Lambda_0^{\; 1} \frac{\partial}{\partial x^1}
\label{eq:time_comp_1}
\end{equation}

\begin{equation}
\frac{\partial}{\partial x'^0} = \gamma_0 \frac{\partial}{\partial x^0} + \gamma_0 \beta_0 \frac{\partial}{\partial x^1}
\label{eq:time_comp_2}
\end{equation}

Since $\partial_0 = (1/c)\partial_t$ and $\partial_1 = \partial_x$:

\begin{equation}
\frac{1}{c}\frac{\partial}{\partial t'} = \gamma_0 \frac{1}{c}\frac{\partial}{\partial t} + \gamma_0 \beta_0 \frac{\partial}{\partial x}
\label{eq:time_comp_3}
\end{equation}

\begin{equation}
\frac{\partial}{\partial t'} = \gamma_0 \left( \frac{\partial}{\partial t} + v_0 \frac{\partial}{\partial x} \right).
\label{eq:time_transform}
\end{equation}

Collecting all components:

\begin{subequations}
\label{eq:partial_summary}
\begin{align}
\frac{\partial}{\partial t'} &= \gamma_0 \left( \frac{\partial}{\partial t} + v_0 \frac{\partial}{\partial x} \right), \label{eq:partial_t} \\
\frac{\partial}{\partial x'} &= \gamma_0 \left( \frac{\partial}{\partial x} + \frac{v_0}{c^2} \frac{\partial}{\partial t} \right), \label{eq:partial_x} \\
\frac{\partial}{\partial y'} &= \frac{\partial}{\partial y}, \label{eq:partial_y} \\
\frac{\partial}{\partial z'} &= \frac{\partial}{\partial z}. \label{eq:partial_z}
\end{align}
\end{subequations}

\textbf{c)} The electromagnetic field tensor transforms as a rank-2 tensor, and the transformation laws are:

\begin{subequations}
\begin{align}
F'^{ik} &= \Lambda^i_{\; m} \Lambda^k_{\; n} F^{mn}, \label{eq:tensor_transform_a} \\
F'_{ik} &= \Lambda_i^{\; m} \Lambda_k^{\; n} F_{mn}, \label{eq:tensor_transform_b}
\end{align}
\end{subequations}
for the contravariant and covariant forms, respectively
After applying the Lorentz transformation, the transformed field tensor in the moving frame is:

\begin{equation}
F'^{ik} = 
\begin{pmatrix}
0 & -E_x'/c & -E_y'/c & -E_z'/c \\
E_x'/c & 0 & -B_z' & B_y' \\
E_y'/c & B_z' & 0 & -B_x' \\
E_z'/c & -B_y' & B_x' & 0
\end{pmatrix}
\label{eq:F_tensor_transformed}
\end{equation}
with the transformed field components given by:
\begin{equation}
\begin{aligned}
E_x' &= E_x, \\
E_y' &= \gamma_0 (E_y - v_0 B_z), \\
E_z' &= \gamma_0 (E_z + v_0 B_y), \\
B_x' &= B_x, \\
B_y' &= \gamma_0 \left( B_y + \frac{v_0}{c^2} E_z \right), \\
B_z' &= \gamma_0 \left( B_z - \frac{v_0}{c^2} E_y \right).
\end{aligned}
\label{eq:field_transform}
\end{equation}

Inserting the transformed expressions for the four-velocity, 
the partial derivative with respect to the four-position, 
and the electromagnetic field tensor—given in Eqs.~\eqref{eq:4velocity_transform}, \eqref{eq:partial_transform}, and \eqref{eq:tensor_transform_a}--\eqref{eq:tensor_transform_b}, 
respectively—into Eq.~\eqref{eq:4force} yields the Lorentz-transformed LL force:

\textbf{a)} The first term of Eq.~\eqref{eq:4force} is:

\begin{equation}
T_1 = -\frac{e}{mc^2}
\frac{\partial F^{ik}}{\partial x^l}
u_{k}u^l.
\label{eq:term1_original}
\end{equation}

Substituting the transformed quantities:

\begin{equation}
T'_1 = -\frac{e}{mc^2}
\left( \frac{\partial}{\partial x'^l} F'^{ik} \right)
u'_k u'^l.
\label{eq:term1_transformed_1}
\end{equation}

Using Eqs.\eqref{eq:4velocity_transform}, \eqref{eq:partial_transform}, and \eqref{eq:tensor_transform_a}--\eqref{eq:tensor_transform_b}

\begin{equation}
T'_1 = -\frac{e}{mc^2}
\Lambda^i_{\; m} \Lambda^k_{\; n} \Lambda_l^{\; p} \Lambda^l_{\; q}
\frac{\partial F^{mn}}{\partial x^p}
u_k u^q.
\label{eq:term1_transformed_2}
\end{equation}

Using the orthogonality relation $\Lambda_l^{\; p} \Lambda^l_{\; q} = \delta^p_q$:

\begin{equation}
T'_1 = -\frac{e}{mc^2}
\Lambda^i_{\; m} \Lambda^k_{\; n}
\frac{\partial F^{mn}}{\partial x^p}
u_k u^p.
\label{eq:term1_transformed_final}
\end{equation}

\textbf{b)}  The second term of Eq.~\eqref{eq:4force} is:

\begin{equation}
T_2 = -\frac{e^2}{m^2c^4}
F^{ik}F_{lk}u^l.
\label{eq:term2_original}
\end{equation}

Substituting the transformed quantities:

\begin{equation}
T'_2 = -\frac{e^2}{m^2c^4}
F'^{ik} F'_{lk} u'^l.
\label{eq:term2_transformed_1}
\end{equation}

Using $F'^{ik} = \Lambda^i_{\; m} \Lambda^k_{\; n} F^{mn}$, $F'_{lk} = \Lambda_l^{\; p} \Lambda_k^{\; q} F_{pq}$, and $u'^l = \Lambda^l_{\; r} u^r$:

\begin{equation}
T'_2 = -\frac{e^2}{m^2c^4}
\left( \Lambda^i_{\; m} \Lambda^k_{\; n} F^{mn} \right)
\left( \Lambda_l^{\; p} \Lambda_k^{\; q} F_{pq} \right)
\left( \Lambda^l_{\; r} u^r \right).
\label{eq:term2_transformed_2}
\end{equation}

Using the orthogonality relations $\Lambda^k_{\; n} \Lambda_k^{\; q} = \delta_n^q$ and $\Lambda_l^{\; p} \Lambda^l_{\; r} = \delta_r^p$:

\begin{equation}
T'_2 = -\frac{e^2}{m^2c^4}
\Lambda^i_{\; m}
F^{mn} F_{pn} u^p.
\label{eq:term2_transformed_final}
\end{equation}

\textbf{c)}  The third term of Eq.~\eqref{eq:4force} is:

\begin{equation}
T_3 = +\frac{e^2}{m^2c^4}
(F_{lk}u^k)
(F^{lm}u_{m})u^i.
\label{eq:term3_original}
\end{equation}

Substituting the transformed quantities:

\begin{equation}
T'_3 = +\frac{e^2}{m^2c^4}
\left( F'_{lk} u'^k \right)
\left( F'^{lm} u'_m \right)
u'^i.
\label{eq:term3_transformed_1}
\end{equation}

Using the transformed field tensor components and 4-velocity:

\begin{equation}
T'_3 = +\frac{e^2}{m^2c^4}
\left( \Lambda_l^{\; p} \Lambda_k^{\; q} F_{pq} \right)
\left( \Lambda^k_{\; r} u^r \right)
\left( \Lambda^l_{\; s} \Lambda^m_{\; t} F^{st} \right)
\left( \Lambda_m^{\; u} u_u \right)
\left( \Lambda^i_{\; v} u^v \right).
\label{eq:term3_transformed_2}
\end{equation}

Using the orthogonality relations $\Lambda_k^{\; q} \Lambda^k_{\; r} = \delta_r^q$, $\Lambda_l^{\; p} \Lambda^l_{\; s} = \delta_s^p$, and $\Lambda^m_{\; t} \Lambda_m^{\; u} = \delta_t^u$:

\begin{equation}
T'_3 = +\frac{e^2}{m^2c^4}
\Lambda^i_{\; v}
\left( F_{pq} u^q \right)
\left( F^{pt} u_t \right)
u^v.
\label{eq:term3_transformed_final}
\end{equation}

Collecting all transformed terms from Eqs.~\eqref{eq:term1_transformed_final}, \eqref{eq:term2_transformed_final}, and \eqref{eq:term3_transformed_final}:

\begin{multline}
g'^i = \frac{2}{3}\frac{e^2}{c}
\Biggl[
-\frac{e}{mc^2}
\Lambda^i_{\; m} \Lambda^k_{\; n}
\frac{\partial F^{mn}}{\partial x^p}
u_k u^p \\
-\frac{e^2}{m^2c^4}
\Lambda^i_{\; m}
F^{mn} F_{pn} u^p
+\frac{e^2}{m^2c^4}
\Lambda^i_{\; v}
\left( F_{pq} u^q \right)
\left( F^{pt} u_t \right)
u^v
\Biggr].
\label{eq:g_prime_combined}
\end{multline}

Factoring out the Lorentz transformation matrix $\Lambda^i_{\; j}$:

\begin{multline}
g'^i = \Lambda^i_{\; j}
\frac{2}{3}\frac{e^2}{c}
\Biggl[
-\frac{e}{mc^2}
\frac{\partial F^{jk}}{\partial x^p}
u_k u^p \\
-\frac{e^2}{m^2c^4}
F^{jk} F_{pk} u^p
+\frac{e^2}{m^2c^4}
\left( F_{pq} u^q \right)
\left( F^{jp} u_p \right)
\Biggr].
\label{eq:g_prime_factored}
\end{multline}
This result is in full agreement with the contravariant 4-vector transformation law of Eq.~\eqref{eq:4force_transform}.

\vspace{5cm}

\newpage
\bibliographystyle{apsrev4-2} 
\bibliography{journals
}

\end{document}